\documentclass[aps,pra,twocolumn,showpacs,floatfix,superscriptaddress]{revtex4-1}
\usepackage[utf8]{inputenc}
\usepackage[T1]{fontenc}
\usepackage[english]{babel}
\usepackage{float}
\usepackage{color}
\usepackage{graphicx}
\usepackage{amsmath}
\usepackage{amssymb}
\usepackage{bbm}
\usepackage{indentfirst}
\usepackage{dcolumn}
\usepackage{soul}
\usepackage[caption=false]{subfig}

\begin{document}
\title{Domain wall profiles in Co/Ir$_{n}$/Pt(111) ultrathin films: influence of the Dzyaloshinskii-Moriya interaction}

\author{Gy. J. Vida}
\affiliation{Department of Theoretical Physics, Budapest University of Technology and Economics, Budafoki \'{u}t 8, H-1111 Budapest, Hungary}
\author{E. Simon}
\affiliation{Department of Theoretical Physics, Budapest University of Technology and Economics, Budafoki \'{u}t 8, H-1111 Budapest, Hungary}
\author{L. R\'{o}zsa}
\affiliation{Institute for Solid State Physics and Optics, Wigner Research Centre for Physics, Hungarian Academy of Sciences,
P.O. Box 49, H-1525 Budapest, Hungary}
\author{K. Palot\'{a}s}
\affiliation{Department of Theoretical Physics, Budapest University of Technology and Economics, Budafoki \'{u}t 8, H-1111 Budapest, Hungary}
\affiliation{Department of Complex Physical Systems, Institute of Physics, Slovak Academy of Sciences, SK-84511 Bratislava, Slovakia}
\author{L. Szunyogh}
\affiliation{Department of Theoretical Physics, Budapest University of Technology and Economics, Budafoki \'{u}t 8, H-1111 Budapest, Hungary}
\affiliation{MTA-BME Condensed Matter Research Group, Budapest University of Technology and Economics, Budafoki \'{u}t 8, H-1111 Budapest, Hungary}

\begin{abstract}
We perform a study of domain walls in Co/Ir$_{n}$/Pt(111) ($n=0,\dots,6$) films by a combined approach of first-principles calculations and spin dynamics simulations. 
We determine the tensorial exchange interactions and the magnetic anisotropies for the Co overlayer in both FCC and HCP geometries, depending on the number of Ir buffer layers. We find strong ferromagnetic nearest-neighbor isotropic exchange interactions between the Co atoms and an out-of-plane magnetic anisotropy for the films in FCC geometry. Our simulations show that the magnetic domain walls are of N\'{e}el type, and their rotational sense (chirality) is changed upon the insertion of an Ir buffer layer as compared to the pristine Co/Pt(111) system. Our spin dynamics simulations indicate a twisting of the spins with respect to the planar domain wall profile on the triangular lattice. We discuss this domain wall twisting using symmetry arguments and in terms of an appropriate micromagnetic continuum model considering extra energy terms compared to the available literature.
\end{abstract}

\maketitle

\section{Introduction}

Effective spin models are widely used to investigate the magnetic properties of solids. The breaking of inversion symmetry in noncentrosymmetric crystals, at surfaces or interfaces and the presence of the spin--orbit coupling lead to the appearance of an anisotropic exchange term beyond the isotropic Heisenberg interaction, which is known as the Dzyaloshinskii-Moriya (DM) interaction\cite{Dzyaloshinsky, Moriya}. In collinear ferromagnetic systems, this type of interaction provides domain walls (DWs) with a chiral character\cite{Thiaville-DW,Ryu-spintorque,Chen-DW,Chen-PRL,Bergmann-JPCM,Tetienne2015}, plays a key role in DW dynamics\cite{Moore-APL,PhysRevB.88.214401,Emori-DW, PhysRevB.91.014433}, and leads to the stabilization of isolated chiral skyrmions\cite{Bogdanov,Nagaosa,Rohart-PRB,Romming-Science}. It may also cause the formation of noncollinear magnetic states\cite{Bergmann-JPCM,PhysRevLett.102.067207} such as spin spirals\cite{Bode-Nature,Meckler-PRL} and condensated skyrmionic phases\cite{Rossler,Muhlbauer-Science,Yu-Nature,Heinze-Science,Romming-Science,Esimon-PRB}. Furthermore, the DM interaction induces an asymmetry in the spin wave spectrum of thin ferromagnetic films\cite{Udvardi,Cortes-Ortuno}. Based on this asymmetry, recently extensive experimental efforts have been directed towards the measurement of the interfacial DM interaction by using inelastic light scattering\cite{PhysRevB.91.180405,Cho2015,Kim-ApplPhys}, highly resolved spin-polarized electron energy loss\cite{PhysRevLett.104.137203} or propagating spin wave spectroscopy\cite{acs.nanolett.5b02732}.

The current-driven motion of domain walls is mainly investigated in ultrathin films and multilayers, paving the way for future applications in spintronic and logic devices\cite{Allwood,Parkin}. In these systems, heavy nonmagnetic elements provide the strong spin--orbit coupling necessary for the appearance of the DM interaction in the adjacent magnetic layers. Using the micromagnetic energy functional determined by Dzyaloshinskii\cite{Dzyaloshinsky2}, it has been demonstrated\cite{Bogdanov2,Heide-PRB,Thiaville-DW} that the DM interaction prefers a cycloidal or N\'{e}el-type rotation of spins within a domain wall in the $C_{n\textrm{v}}$ symmetry class to which the majority of these systems belong. The rotational plane of domain walls is determined by the competition between the DM interaction and the magnetostatic dipolar interaction preferring a helical or Bloch-type rotation\cite{Thiaville-DW}, while the right- or left-handed chirality is determined by the sign of the DM interaction. Recently a significant research effort has been devoted to examine the connection between the nonmagnetic material composition and the sign and magnitude of the DM interaction, both based on experimental observations\cite{Ryu-spintorque,Ryu2,Chen-DW,acs.nanolett.5b02732,PhysRevB.90.020402} and first-principles electronic structure calculations\cite{PhysRevLett.115.267210}. In Ref.~\cite{PhysRevB.90.020402}, Pt/Co/Pt and Pt/Co/Ir/Pt multilayers with different Ir thicknesses were studied using a field-driven domain wall creep-based method. It was demonstrated that due to the insertion of the Ir layer, the chirality of the N\'{e}el wall reversed from right-handed to left-handed, which was attributed to the sign change of the effective DM interaction.

Motivated by the experimental study in Ref.~\cite{PhysRevB.90.020402}, in this work we investigate the magnetic properties of Co/Pt(111) and Co/Ir$_{n}$/Pt(111) ($n=1,\dots,6$) films. We use first-principles electronic structure calculation methods to determine the parameters in a spin model where the coupling between the spins is described by an exchange interaction tensor\cite{rtm-Udvardi}, and perform atomistic spin dynamics calculations in order to determine the domain wall profiles. We give direct evidence of the relationship between the homochirality of the N\'{e}el DWs and the calculated DM vectors, confirming the reversal of the DW chirality by inserting the Ir layer between the Co monolayer and the Pt substrate. Moreover, we observe that the presence of the Ir layers weakens the ferromagnetic exchange coupling between the neighboring Co atoms and increases the magnetic anisotropy, thus the DWs become more narrow. We also observe a small twisting of the spins in the DW, leading to a non-coplanar DW profile. Using symmetry arguments we attribute the appearance of this twisting to the out-of-plane components of the DM vectors and a specific term appearing in the symmetric off-diagonal part of the interaction tensors. We explain how the appearance of the twisting depends on the DW normal vector direction, and construct an appropriate micromagnetic continuum model, where the direction and shape of the twisting depends on the coefficients corresponding to the out-of-plane component of the DM vector and the symmetric off-diagonal interaction.

\section{Computational method\label{sec2}}

We performed self-consistent electronic structure calculations for the Co/Ir$_{n}$/Pt(111) ($n=0,\dots,6$) ultrathin films in terms of relativistic screened Korringa--Kohn--Rostoker (SKKR) method\cite{Szunyogh,Zeller}. We used the local spin density approximation parametrized by Vosko et al.\cite{vosko} and the atomic sphere approximation with an angular momentum cutoff of $l_{\textrm{max}}=2$. The system consisted of $10-n$ Pt and $n$ Ir atomic layers ($n=0,\dots,6$), a Co monolayer and $4$ monolayers of vacuum (empty spheres) between the semi-infinite Pt substrate and semi-infinite vacuum. For modeling the geometry of the thin films we used the value $a_{\textrm{2D}}=2.774\,\textrm{\AA}$ for the in-plane lattice constant of the Pt$(111)$ surface, and we optimized the distance between the layers in terms of VASP calculations\cite{Kresse199615, Kresse-PRB, Hafner} for both FCC and HCP stackings of the Co overlayer. We found an inward relaxation of the Co overlayer between 11 and 14\% relative to the Pt-Pt bulk interlayer distance, depending on the number of Ir buffer layers. These inward layer relaxations were used in the self-consistent SKKR calculations, and the Wigner--Seitz radii of the Co, Ir, and top Pt layers were modified according to the relaxations.

We described the localized magnetic moments in the Co layer in terms of a generalized classical Heisenberg model of the form
\begin{equation}
\mathcal{H}=\frac{1}{2}\sum_{ij}\vec{s}_{i}\mathbf{J}_{ij}\vec{s}_{j}+\sum_{i}\vec{s}_{i}\mathbf{K}\vec{s}_{i},
\label{eqn1}
\end{equation}  
where $\vec{s}_{i}$ denotes the spin unit vector at site $i$, $\mathbf{J}_{ij}$ is the exchange coupling tensor\cite{rtm-Udvardi}, and $\mathbf{K}$ is the on-site anisotropy matrix. The corresponding parameters of the spin model in Eq.~(\ref{eqn1}) were determined by combining the SKKR method with the relativistic torque technique\cite{rtm-Udvardi, rtm-Ebert}, based on calculating the energy costs of infinitesimal rotations around ferromagnetic states oriented along different crystallographic directions. We considered the out-of-plane ferromagnetic state and orientations along three nonparallel in-plane nearest-neighbor directions, sufficient for reproducing the $C_{3\textrm{v}}$ symmetry of the system in the interaction tensors. The energy integrals were performed by sampling 16 points on a semicircle contour in the upper complex semi-plane. We have calculated the interactions with neighbors within a radius of $5a_{\textrm{2D}}$, for a total of 90 neighbors.

The interaction tensor $\mathbf{J}_{ij}$ may be decomposed as
\begin{equation}
\mathbf{J}_{ij}=\frac{1}{3} \textrm{Tr}\mathbf{J}_{ij}\mathbf{I}+\frac{1}{2}\left(\mathbf{J}_{ij}-\mathbf{J}^{T}_{ij}\right)+\left[\frac{1}{2}\left(\mathbf{J}_{ij}+\mathbf{J}^{T}_{ij}\right)-\frac{1}{3} \textrm{Tr}\mathbf{J}_{ij}\mathbf{I}\right]  ,
\end{equation}
i.e. into an isotropic, an antisymmetric and a traceless symmetric part. The first term represents the scalar Heisenberg couplings between the magnetic moments, with $J_{ij}=\frac{1}{3} \textrm{Tr}\mathbf{J}_{ij}$. The three components of the antisymmetric part of the exchange tensor can be identified with the DM vector $\vec{D}_{ij}$, defined as
\begin{equation}
\vec{s}_{i}\frac{1}{2}\left(\mathbf{J}_{ij}-\mathbf{J}^{T}_{ij}\right)\vec{s}_{j}=\vec{D}_{ij}\left(\vec{s}_{i}\times\vec{s}_{j}\right).\label{eqn3}
\end{equation}
The traceless symmetric part contains five components in the general case; its diagonal terms induce an energy difference between the uniformly magnetized states along the out-of-plane ($z$) and in-plane ($x$) directions, $\Delta J=\frac{1}{2}
%\sum_{\boldsymbol{R}_{i}-\boldsymbol{R}_{j}}
\sum_{j}\left(J_{ij}^{xx}-J_{ij}^{zz}\right)$, which we will refer to as the two-site magnetic anisotropy. In the $C_{3\textrm{v}}$ symmetry class, the on-site anisotropy tensor may be described by a single parameter,
\begin{equation}
\vec{s}_{i}\mathbf{K}\vec{s}_{i}=-K\left(s^{z}_{i}\right)^{2}.\label{eqn4}
\end{equation}
The total magnetic anisotropy energy (MAE) of the system can be expressed as a sum of the on-site and two-site contributions, $\textrm{MAE}=K+\Delta J$.

In order to determine the equilibrium DW profile, we performed spin dynamics simulations by numerically solving the deterministic Landau--Lifshitz--Gilbert equation\cite{Landau,Gilbert},
\begin{equation}
\partial_{t}\vec{s}_{i}=-\gamma'\vec{s}_{i}\times\vec{B}_{i}^{\textrm{eff}}-\alpha\gamma'\vec{s}_{i}\times\left(\vec{s}_{i}\times\vec{B}_{i}^{\textrm{eff}}\right),\label{eqn5}
\end{equation}
with $\alpha$ the Gilbert damping parameter and $\gamma'=\frac{\alpha}{1+\alpha^{2}}\gamma$, where $\gamma=\frac{ge}{2m}$ is the gyromagnetic ratio ($g$, $e$ and $m$ standing for the electron $g$ factor, absolute charge, and mass). The spin model parameters from Eq.~(\ref{eqn1}) enter into the effective field $\vec{B}_{i}^{\textrm{eff}}=-\frac{1}{M}\frac{\partial H}{\partial \vec{s}_{i}}$, which governs the time evolution of the spins. The spin magnetic moment $M$ was determined from the electronic structure calculations, taking values between $1.9$ and $2.1\,\mu_{\textrm{B}}$ depending on the number of Ir layers and FCC or HCP stacking.

\section{Results and discussion}

\subsection{Spin model parameters\label{sec3a}}

\begin{figure}
\centering
\includegraphics[width=1.0\columnwidth]{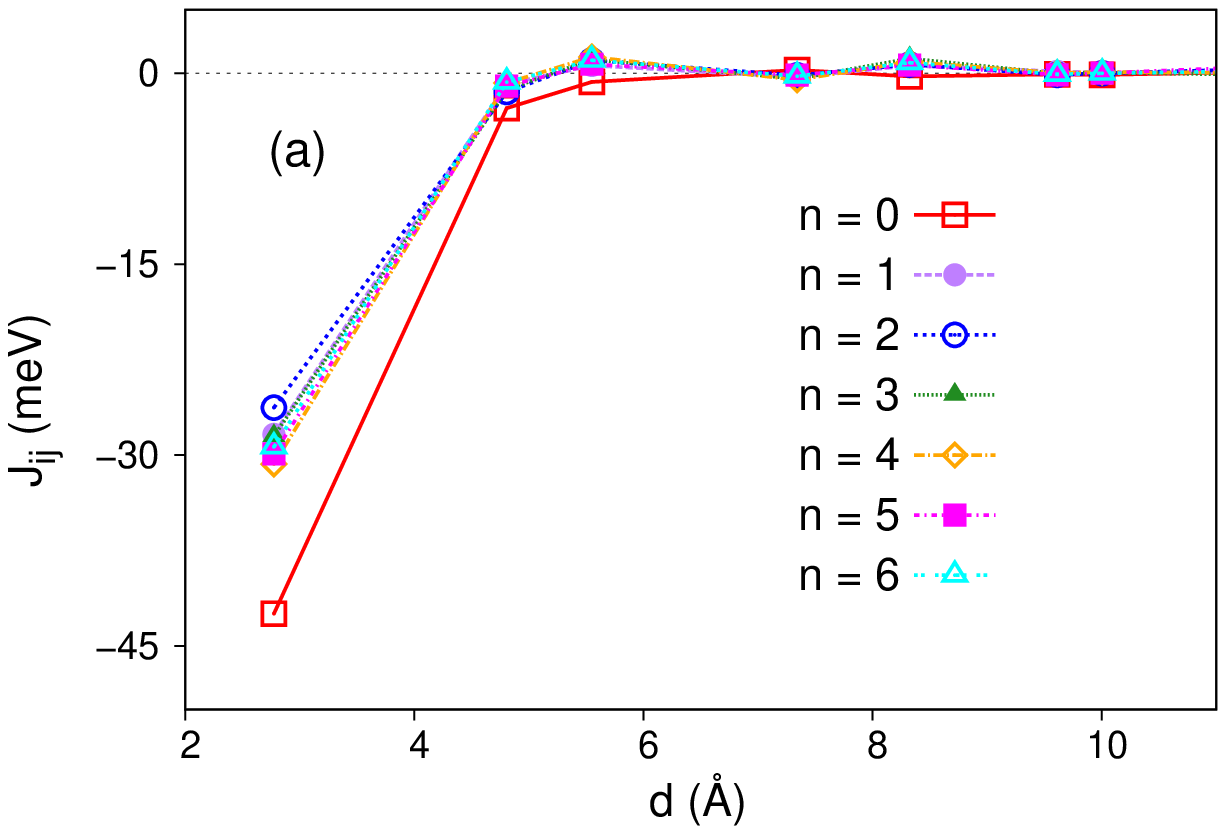}\\
\includegraphics[width=1.0\columnwidth]{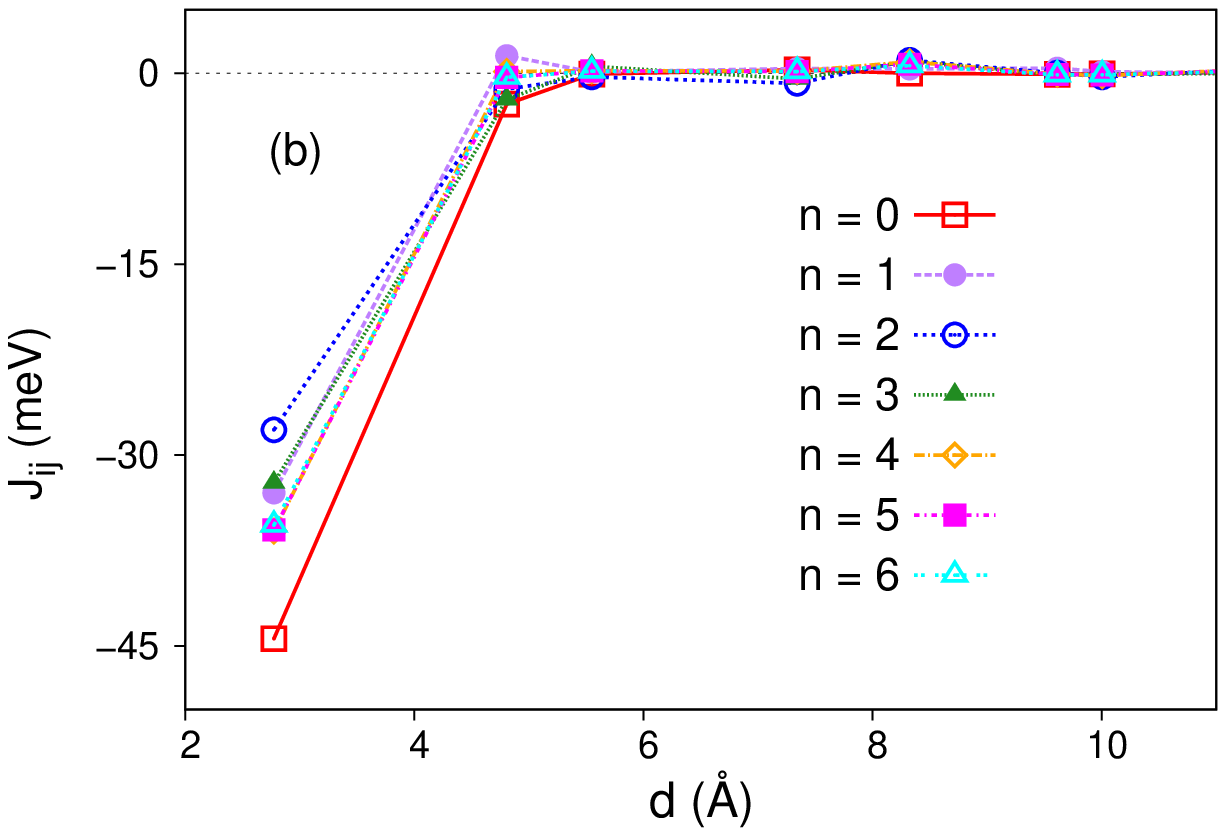}
\caption{(Color online) Calculated Co-Co isotropic exchange parameters as a function of the interatomic distance for (a) FCC and (b) HCP stacking geometries of the Co overlayer, for different numbers of Ir buffer layers.}
\label{jij}
\end{figure}

Figure~\ref{jij} displays the calculated isotropic exchange constants $J_{ij}$ between the Co atoms as a function of the interatomic distance for FCC and HCP stackings, for different numbers of Ir buffer layers.
Because of the sign convention of Eq.~(\ref{eqn1}), the negative sign of the isotropic exchange parameter means ferromagnetic (FM) coupling, while the positive sign refers to antiferromagnetic (AFM) interaction. For both types of stacking, the nearest-neighbor (NN) interactions are strongly FM; for larger interatomic distances the interactions are mostly negligible due to the rapid decay. The NN coupling is the strongest for Co/Pt(111), $-42.5\,\textrm{meV}$ and $-44.4\,\textrm{meV}$ in FCC and HCP stacking, respectively. The presence of Ir layers considerably reduces the NN coupling, which is almost independent of the number of buffer layers for FCC stacking (between $-27$ and $-30\,\textrm{meV}$), while in the case of the HCP stacking this range is somewhat wider (between $-27$ and $-35\,\textrm{meV}$).

We have summarized the on-site, two-site, and total magnetic anisotropies for the Co monolayer in Table~\ref{anisotropies}, considering both types of stacking.
\begin{table}
\centering
\begin{ruledtabular}
\begin{tabular}{c   r r r r r r}
 & \multicolumn{3}{c}{FCC} &  \multicolumn{3}{c}{HCP}\\ 
$n$ & $K$ & $ \Delta J$ & MAE & $K$ & $ \Delta J$ & MAE \\ 
\hline
0 & 0.40 & -0.20 & 0.20 &  0.46 &  0.10 &  0.57  \\ 
1 & 1.02 &  1.63 & 2.64 &  1.17 &  1.41 &  2.58  \\ 
2 & 0.77 &  1.01 & 1.77 &  0.31 & -0.38 & -0.06 \\  
3 & 0.31 &  0.60 & 0.91 & -0.06 & -0.87 & -0.93 \\  
4 & 0.82 &  0.56 & 1.39 &  0.70 &  0.02 &   0.72\\  
5 & 0.87 &  1.66 & 2.53 &  0.81 &  0.36 &   1.17\\  
6 & 0.75 &  1.71 & 2.46 &  0.57 &  0.25 &   0.82\\ 
\end{tabular}
\end{ruledtabular} 
\caption{Calculated on-site ($K$), two-site ($\Delta J$) and total ($\textrm{MAE}=K+\Delta J$) magnetic anisotropies for the Co monolayer as a function of the number of Ir buffer layers in Co/Ir$_n$/Pt(111). All data are normalized to a single Co atom and are given in meV units. 
The positive sign refers to the $z$ (out-of-plane) easy axis, while the negative sign means easy-plane anisotropy.}
\label{anisotropies}
\end{table}
With our definition the positive sign of the on-site and two-site magnetic anisotropies corresponds to an easy axis along the out-of-plane ($z$) direction. It can be seen from Table~\ref{anisotropies} that most of the samples have an out-of-plane easy axis. For FCC stacking, the Ir buffer layer clearly enhances the magnetic anisotropy, which seems to saturate at around $2.5~\textrm{meV}$ for larger $n$. For HCP stacking, an Ir monolayer also remarkably increases the perpendicular MAE. In the case of two and three Ir atomic layers we, however, observe easy-plane anisotropy, while for thicker Ir layers it is again of easy-axis type. This oscillation of the sign of the MAE is similar to the effect recently found in Mn/W$_{m}$/Co$_{n}$/W$(001)$ multilayers\cite{PhysRevLett.116.177202}, and can most likely be attributed to interface-induced Friedel oscillations.

\begin{figure}
\centering
\includegraphics[width=1.0\columnwidth]{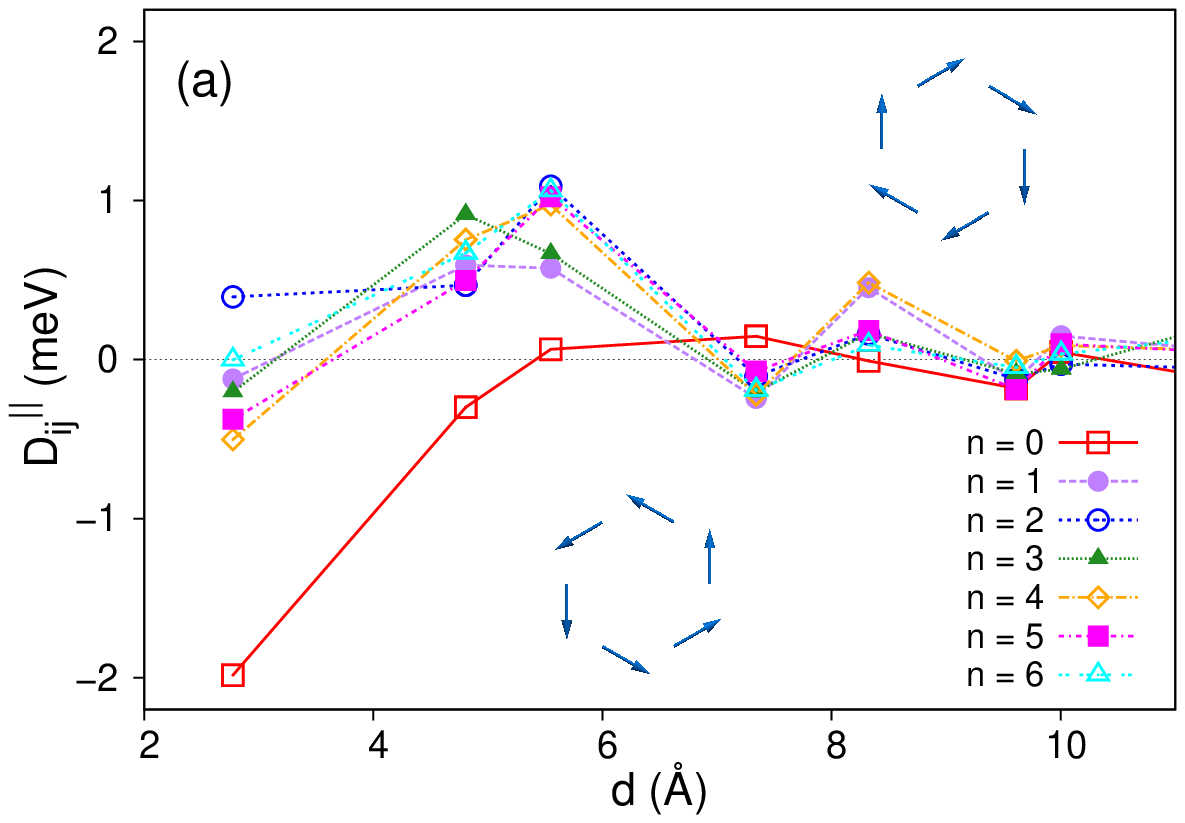}\\
\includegraphics[width=1.0\columnwidth]{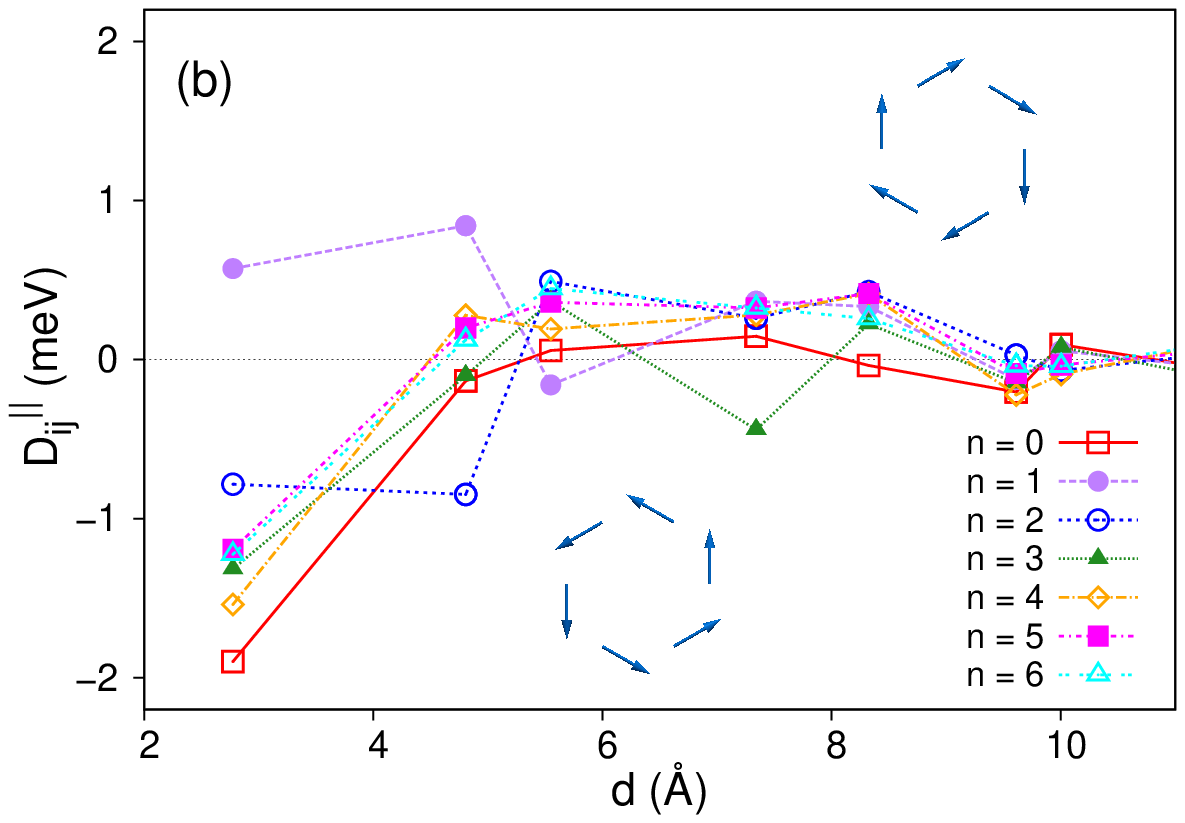}
\caption{(Color online) In-plane components of the DM vectors,  $D_{ij}^{\Vert}$ , as a function of distance between the Co atoms and the thickness of the Ir buffer layers in case of (a) FCC and (b) HCP stacking geometries. The insets illustrate the rotational direction of the DM vectors encoded in the sign of $D_{ij}^{\Vert}$.}
\label{in-plane-DM}
\end{figure}

Next we investigate the in-plane component of the DM vectors, $D_{ij}^{\Vert}$, between the Co atoms for FCC and HCP stacking geometries, since this component is related to the strength of the scalar DM interaction in micromagnetic models, see Ref.~\cite{PhysRevLett.115.267210} and Appendix~\ref{secS1}. The sign of $D_{ij}^{\Vert}$ corresponds to the rotational direction of the DM vectors in a given shell of neighbors as illustrated in Fig.~\ref{in-plane-DM}. In case of Co/Pt(111), $D_{ij}^{\Vert}$ for the NN`s with the value of $-1.98\,\textrm{meV}$ for the FCC stacking and $-1.89\,\textrm{meV}$ for the HCP stacking are the most significant, and they rotate in counter-clockwise direction. For Co/Ir$_{n}$/Pt(111) ($n=1,\dots,6$) layers with FCC stacking geometry the magnitude of the NN in-plane DM vectors is much smaller than for Co/Pt(111), and 
$D_{ij}^{\Vert}$ for the second and third neighbors dominate with a clockwise rotational direction, denoted by a positive sign in Fig.~\ref{in-plane-DM}(a).  This clearly implies a sign change of the effective scalar DM interaction when adding Ir layers between the Pt and Co layers, similarly to the recent experimental\cite{PhysRevB.90.020402} as well as theoretical findings\cite{PhysRevLett.115.267210}. In case of HCP stacking geometry (see Fig.~\ref{in-plane-DM}(b)), for Co/Ir$_{1}$/Pt(111)  the first- and second-neighbor in-plane DM vectors dominate with approximately the same magnitude and rotating in a clock-wise direction, i.e. the sign change of the scalar DM interaction is present. For thicker Ir layers, the  NN $D_{ij}^{\Vert}$ with relatively large magnitude turns to counter-clockwise direction, while the magnitude and direction of $D_{ij}^{\Vert}$ show an oscillating behavior against both the distance between the Co atoms and the number of Ir layers.

From the calculated spin model parameters, it can be concluded that the magnetic anisotropy and the DM vectors strongly depend on the stacking geometry. We attribute the high sensitivity of the interactions induced by spin--orbit coupling to the different hybridization between the electronic states of the Co monolayer and the adjacent Ir layer for the different stackings. Similar effects related to the stacking geometry have been reported experimentally for a Mn monolayer on Ag(111)\cite{PhysRevLett.101.267205} or Fe/Ir(111)\cite{NanoLett.15.3280}, and computationally for Cr/Au(111)\cite{0953-8984-26-43-436001} or for Pd/Fe/Ir(111)\cite{NatComm.5.4030}. Since in the FCC geometry the system is perpendicularly magnetized regardless of the thickness of the Ir buffer layer, which corresponds to the experimental situation\cite{PhysRevB.90.020402}, in the next sections we focus on the domain wall formation in the Co/Pt(111) and Co/Ir$_{n}$/Pt(111) films only in case of FCC stacking.

\subsection{Domain wall formation and chirality\label{sec3b}}

By using the spin model parameters obtained from first principles, we performed spin dynamics simulations for determining the equilibrium DW profiles in the system. We have used a lattice consisting of $N=128\times256$ spins, and set the normal vector of the DW along the $[1\overline{1}0]$ direction connecting two NN sites on a triangular lattice, which will correspond to the $x$ axis of the coordinate system. The perpendicular $[11\overline{2}]$ direction connecting next-nearest neighbors and falling in the symmetry plane of the system will be denoted by $y$. During the simulations we fixed the spins along the $-z$ and $z$ out-of-plane directions at the two edges of the lattice in the $x$ direction, and periodic boundary conditions have been applied along the perpendicular $y$ direction. We have initialized a system in a non-optimized DW configuration, and minimized the energy by numerically solving Eq.~(\ref{eqn5}) with high damping, $\alpha=1$.

\begin{figure}
\centering
\includegraphics[width=1.0\columnwidth]{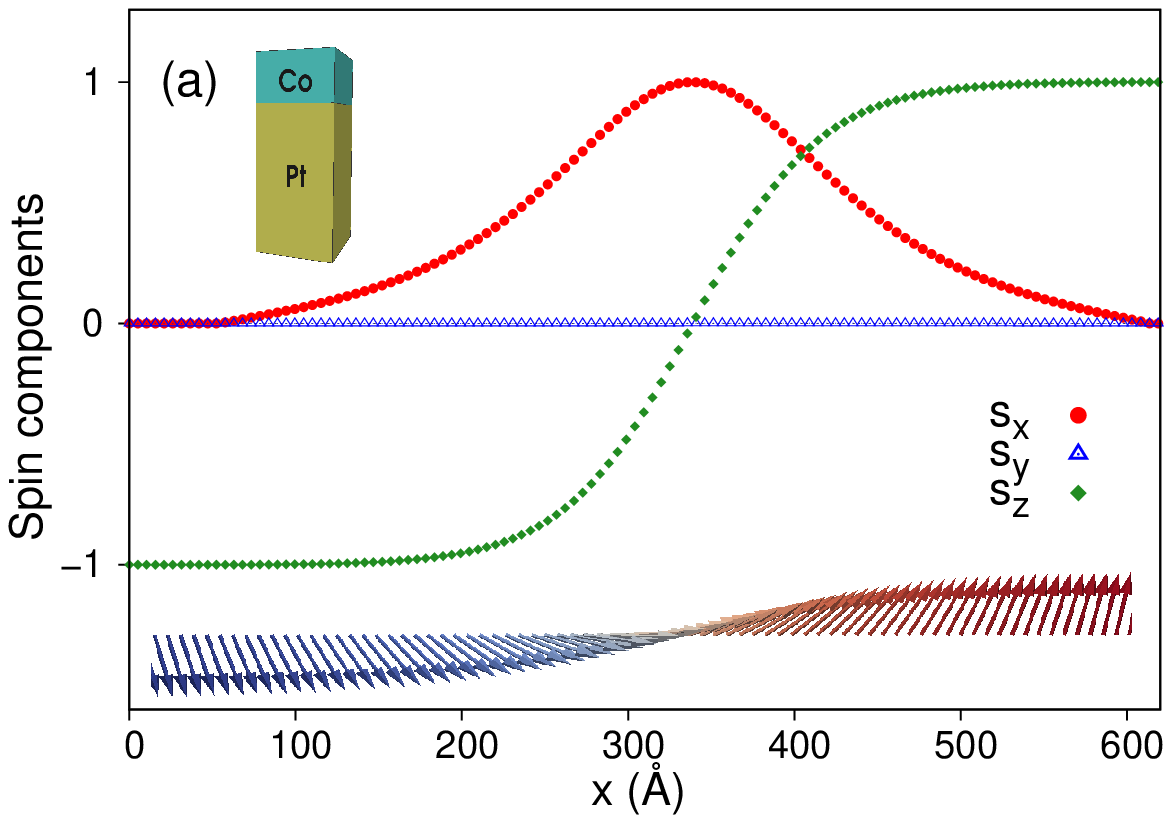}\\
\includegraphics[width=1.0\columnwidth]{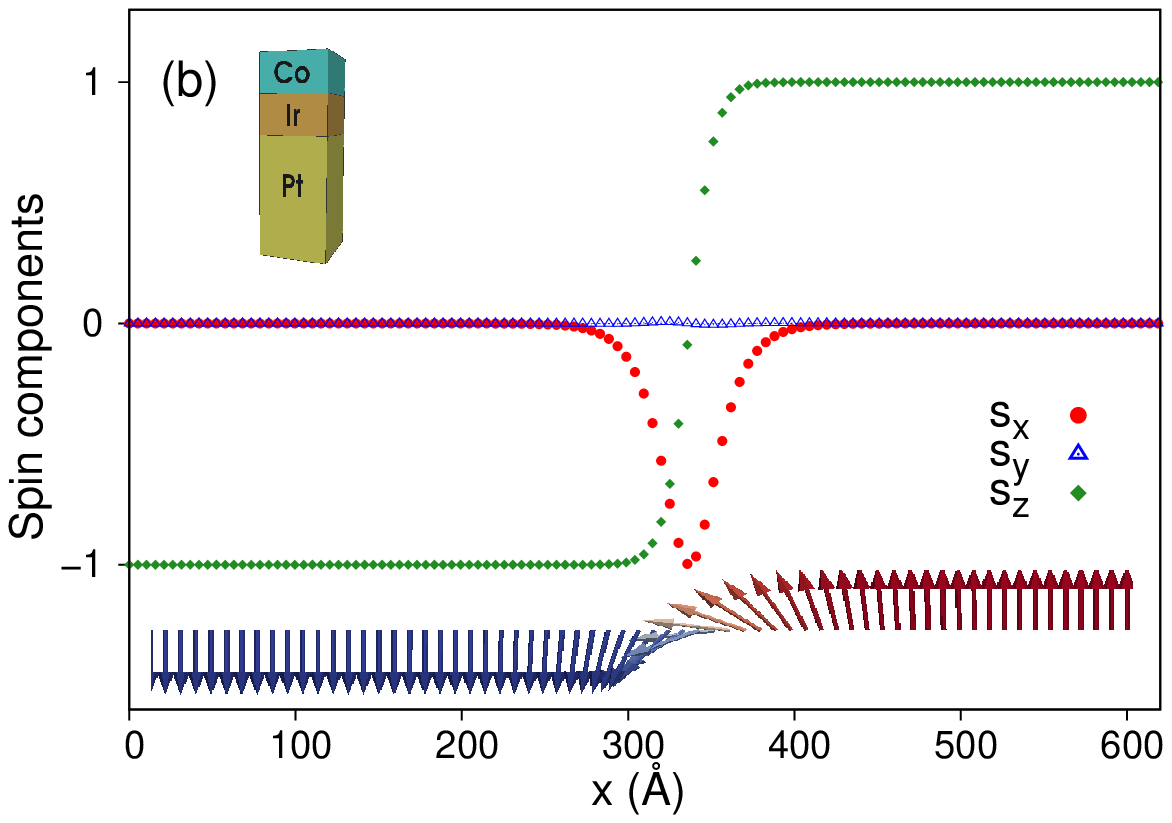}
\caption{(Color online) Domain wall profiles in a DW with normal vector along the $x$ direction, obtained from spin dynamics simulations for (a) Co/Pt(111) and for (b) Co/Ir$_{1}$/Pt(111).} 
\label{spindin}
\end{figure}
%for this reason the perpendicular component of normalized spins (s$_{z}$) changes continuously between up and down directions ($+1$ and $-1$ values of s$_{z}$), while 

The simulated DW profiles are shown for Co/Pt(111) and Co/Ir$_{1}$/Pt(111) films for the FCC stacking of the Co layer in Fig.~\ref{spindin}. The DWs are visibly of N\'{e}el type, since the magnitude of the in-plane component parallel to the propagation direction increases from $s_{x}=0$ at the edge of the sample to $s_{x}=1$ at the center of the wall ($s_{z} =0$), while the $s_{y}$ component remains close to zero, indicating a rotation in the $xz$ plane. It is known from micromagnetic theory\cite{Landau,Heide-PRB} that the DW width is proportional to $\sqrt{\mathcal{A}/\mathcal{K}}$, where the exchange stiffness $\mathcal{A}$ is connected to the Heisenberg exchange $J_{ij}$ in our description, while the anisotropy constant $\mathcal{K}$ corresponds to the MAE. This explains why the width of the DW significantly decreases with the addition of the Ir buffer layer: the NN Heisenberg exchange interaction weakens (see Fig.~\ref{jij}), while the MAE increases (see Table~\ref{anisotropies}).

It can also be seen in Fig.~\ref{spindin} that the rotational sense of the N\'{e}el DW switches from left-handed in Co/Pt(111) to right-handed in Co/Ir$_{1}$/Pt(111), indicated by the sign change of the $s_{x}$ spin component with a fixed sign of $s_{z}$. This is connected to the sign of the in-plane component of the DM vectors in Fig.~\ref{in-plane-DM}: negative and positive signs prefer left- and right-handed rotations, respectively. In the experimental observations of Ref.~\cite{PhysRevB.90.020402}, the rotational sense of the DWs switched from right-handed to left-handed when the Ir buffer layer was introduced between the Co layer and the Pt layer on top of it. The chirality is in agreement with our calculations if we take into account that we have introduced the Ir buffer layer below the Co layer, because swapping the up and down directions also  switches the notion of left- and right-handed rotations\cite{Heide-PRB}.
Our simulations confirmed that by further increasing the number of Ir layers the right-handed DW chirality is preserved, and the DW width is less sensitive to this change. Again these observations are in agreement with the arguments given above and the model parameters discussed in Sec.~\ref{sec3a}.

We also performed the simulations by including the magnetostatic dipolar interaction in Eq.~(\ref{eqn1}). We have included dipolar coupling between neighbors within a radius of $10a_{2\textrm{D}}$, which accounts for about $90\%$ of the total strength of this long-ranged interaction in the considered monolayer system. One effect of the dipolar interaction was decreasing the MAE values listed in Table~\ref{anisotropies} by approximately $0.1\,\textrm{meV}$; however, this does not switch between easy-axis and easy-plane anisotropy in any of the considered cases. Furthermore, it is known from the literature\cite{Bogdanov,Thiaville-DW} that the rotational plane of the DW assumes an intermediate state between Bloch- and N\'{e}el-type rotation if the dipolar interaction is present in the system and the in-plane component of the DM interaction is weaker than a threshold value. However, we have confirmed with simulations that in the considered systems the DM interaction is about ten times stronger than this threshold. Overall, it can be concluded that the dipolar interaction only slightly modifies the DW width in the system, therefore, it can safely be neglected.

\subsection{Atomistic simulations of domain wall twisting\label{sec3c}}

\begin{figure}
\centering
\includegraphics[width=1.00\columnwidth]{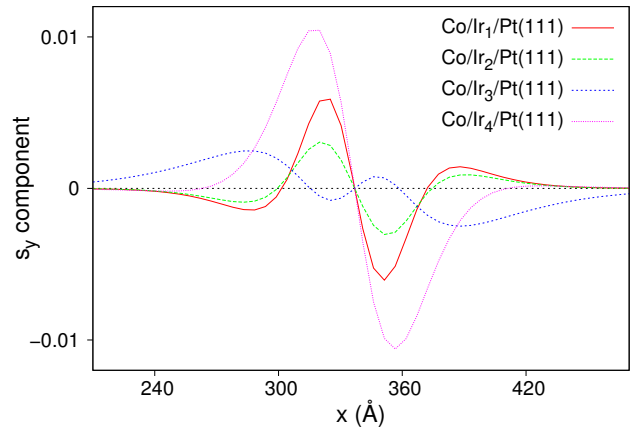}
\caption{(Color online) Twisting of DWs with normal vector along the $x$ direction observable in the $s_{y}$ spin component, obtained from spin dynamics simulations. The figure corresponds to a close-up of Fig.~\ref{spindin}(b) around the center of the wall, for selected numbers of Ir buffer layers.}
\label{deviation}
\end{figure}

From the spin dynamics simulations we observed that the DW profile does not perfectly coincide with a planar N\'{e}el wall if the DW normal vector is along the NN $x$ direction. As demonstrated in Fig.~\ref{deviation}, the $s_{y}$ spin component is also finite within the wall, and analogously to the $s_{z}$ component it changes sign in the middle of the DW. In the following we will refer this modulation of the DW as {\em twisting} of the spins. As illustrated in Fig.~\ref{deviation}, the magnitude and also the exact shape of the twisting depends on the number of Ir buffer layers. However, even in the case of Co/Ir$_{4}$/Pt where the largest twisting occurs, its peak value corresponds to only about 1\% of the total length of the spin vectors.

Note that this twisting is different from the rotation of the complete DW from the N\'{e}el-type towards the Bloch-type, which could occur due to the presence of the dipolar interaction as discussed at the end of Sec.~\ref{sec3b}. In this case the $y$ component of the spin vectors would have a local maximum in the middle of the wall instead of a node. Furthermore, there is apparently no threshold value of the parameters for the occurrence of the twisting, in contrast to the rotation. Finally, we also observed that the twisting completely disappears if the  normal vector of the DW is along the next-nearest-neighbor ($y$) direction, which would not happen in the case of rotation.

By considering symmetry arguments in the atomistic model 
it can be explained why the twisting occurs for DWs with normal vector along the $x$ direction, but not for ones with normal vector along the perpendicular $y$ direction. Considering a N\'{e}el DW with normal vector along the $x$ axis, the system may gain energy from tilting the spins towards the $y$ direction due to the $J_{ij}^{xy}$ and $J_{ij}^{yz}$ components of the $\mathbf{J}_{ij}$ interaction tensor in Eq.~(\ref{eqn1}). As mentioned in Sec.~\ref{sec2}, the antisymmetric part of the tensor may be reformulated in the DM vector $\vec{D}_{ij}$ in Eq.~(\ref{eqn3}). Analogously, from the symmetric part of the off-diagonal components we construct the in-plane vectors
\begin{equation}
\vec{J}_{ij,\textrm{s}}=\left(\frac{1}{2}\left(J_{ij}^{xz}+J_{ij}^{zx}\right),\frac{1}{2}\left(J_{ij}^{yz}+J_{ij}^{zy}\right)\right).\label{eqn6}
\end{equation}

\begin{figure}
\centering
\subfloat[]{\includegraphics[width=0.5\columnwidth]{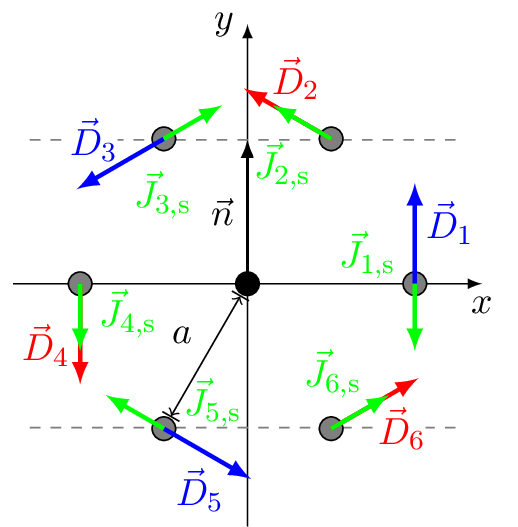}\label{fig:spear}}
\subfloat[]{\includegraphics[width=0.5\columnwidth]{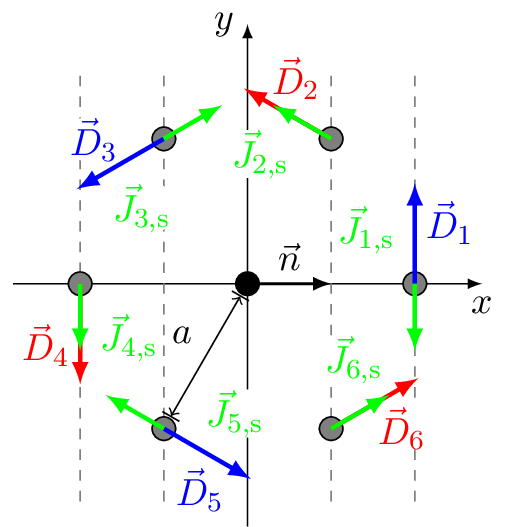}\label{fig:diamond}}
\caption{(Color online) Schematic view of the NN $\vec{D}_{ij}$ and $\vec{J}_{ij,\textrm{s}}$ vectors on a triangular lattice for DWs with normal vectors ($\vec{n}$) along two perpendicular directions. Red and blue colors of the arrows correspond to positive and negative $z$ components of the DM vectors. Staggered lines connect spins with the same orientation in a DW.
\protect\subref{fig:spear} The normal vector is in a mirror plane. The $z$ components of the DM vectors are opposite to each other for equivalent neighbors within the DW, e.g. $D_{1}^{z}=-D_{4}^{z}$, and they do not influence the energy of the system by twisting the DW towards the $y$ direction. Similarly, the sum over $\vec{J}_{ij,\textrm{s}}$ vectors is parallel to the N\'{e}el DW normal vector, which does not induce a twisting. \protect\subref{fig:diamond} The normal vector is perpendicular to the mirror plane. The $z$ components of the corresponding DM vectors appear additively for equivalent neighbors, e.g. $D_{2}^{z}=D_{6}^{z}$, and the sum over $\vec{J}_{ij,\textrm{s}}$ vectors is perpendicular to the $xz$ plane where the spins are oriented in an ideal N\'{e}el DW. Both of these effects lead to a twisting of the DW.}
\label{DMvectors}
\end{figure}

The DM vector transforms as an axial vector, while $\vec{J}_{ij,\textrm{s}}$ transforms as a two-dimensional polar vector under the planar symmetry operations. For the NNs these vectors are illustrated in Fig.~\ref{DMvectors}. Due to the symmetry rules formulated by Moriya\cite{Moriya}, the DM vector must be perpendicular to the lattice vector connecting the NNs; however, it may have an out-of-plane $z$ component, the sign of which is illustrated by red and blue colors in Fig.~\ref{DMvectors}. Note that the out-of-plane component of the DM vectors is allowed specifically for $(111)$ surfaces in cubic systems with $C_{3\textrm{v}}$ symmetry, but it disappears for $(100)$ or $(110)$ surfaces with $C_{4\textrm{v}}$ and $C_{2\textrm{v}}$ symmetries\cite{Crepieux}. Since the $z$ component of the DM vector is connected to the $J_{ij}^{xy}$ tensor element, it may lead to a twisting of the spins. The staggered lines in Fig.~\ref{DMvectors} connect neighbors which are parallel to each other in the DW, for which the out-of-plane components of the DM vector appear additively in the energy expression. If the DW normal vector is in a symmetry plane (along the next-nearest-neighbor $y$ direction, Fig.~\ref{DMvectors}(a)), the $z$ components exactly cancel, and no twisting occurs. However, the $z$ components are of the same sign for equivalent neighbors if the DW normal vector is along the NN $x$ direction (Fig.~\ref{DMvectors}(b)).

Similarly, it can be shown that the $\vec{J}_{ij,\textrm{s}}$ vectors must be perpendicular to the NN lattice vectors. The twisting is caused by their components which is perpendicular to the normal vector of the N\'{e}el DW, corresponding to $J_{ij}^{yz}$ when the normal vector is along the $x$ direction and to $J_{ij}^{xz}$ for the normal vector along the $y$ direction. Similarly to the out-of-plane components of the DM vectors, the components of $\vec{J}_{ij,\textrm{s}}$ cancel for equivalent neighbors if the DW normal vector is along the $y$ direction (Fig.~\ref{DMvectors}(a)), but they may lead to a twisting for DWs with normal vectors along the $x$ direction (Fig.~\ref{DMvectors}(b)).

\subsection{Continuum model of domain wall twisting\label{sec4}}

In order to get further insight  into the formation of the twisting of DWs, we employed a micromagnetic model, where the magnetization is represented by the vector field $\vec{s}(\vec{r})$ with $|\vec{s}|=1$. The appropriate form of the micromagnetic functional containing exchange stiffness, magnetic anisotropy and (in-plane) DM interaction is known from the literature\cite{Dzyaloshinsky2,Heide-PRB}. However, this model has to be extended by terms responsible for the observed twisting in the atomistic model, namely the out-of-plane component of the DM vector and the $\vec{J}_{ij,\textrm{s}}$ vector. The derivation of the appropriate functional in the two-dimensional plane based on symmetry considerations is given in Appendix~\ref{secS1}; here we restrict ourselves to the description of a DW with normal vector along the $x$ direction, which is perpendicular to the mirror plane of the system with $C_{3\textrm{v}}$ symmetry. In this case, the energy expression simplifies to a one-dimensional integral,
\begin{align}
E=& \int \left ( w_{\mathcal{A}}+w_{\mathcal{J}_{\textrm{s}}}+w_{\mathcal{K}}+w_{\mathcal{D}}+w_{\mathcal{D}^{z}}\right)\textrm{d}x\nonumber
\\
=&\int \Big( \mathcal{A} \, \dot{\vec{s}}\left(x\right)^2+ \mathcal{J}_{\textrm{s}}\dot{s}_{y}\left(x\right)\dot{s}_{z}\left(x\right)+ \mathcal{K}s_{z}^{2}\left(x\right)  \nonumber \\
 &+  \mathcal{D} \big[\vec{s}\left(x\right) \times \dot{\vec{s}}\left(x\right)\big]_{y}+\mathcal{D}^{z} \big[\vec{s}\left(x\right) \times \dddot{\vec{s}}\left(x\right)\big]_{z} \Big)\textrm{d}x,\label{functional}
\end{align}
where $\dot{\vec{s}}$ denotes differentiation with respect to the variable $x$. $\mathcal{A}$ corresponds to the exchange stiffness, $\mathcal{D}$ to the linear Lifshitz invariant or DM interaction, and $\mathcal{K}$ to the anisotropy. Note that although the equivalent of the $z$ component of the DM vector $\mathcal{D}^{z}$ is antisymmetric in the spin components as expected, it only appears in a term proportional to the third derivative of the field $\vec{s}$. Finally, the appropriate form $\mathcal{J}_{\textrm{s}}$ of the off-diagonal tensor elements appearing in the vector $\vec{J}_{ij,\textrm{s}}$ is analogous to the exchange stiffness.

\begin{figure}
\includegraphics[width=1.0\columnwidth]{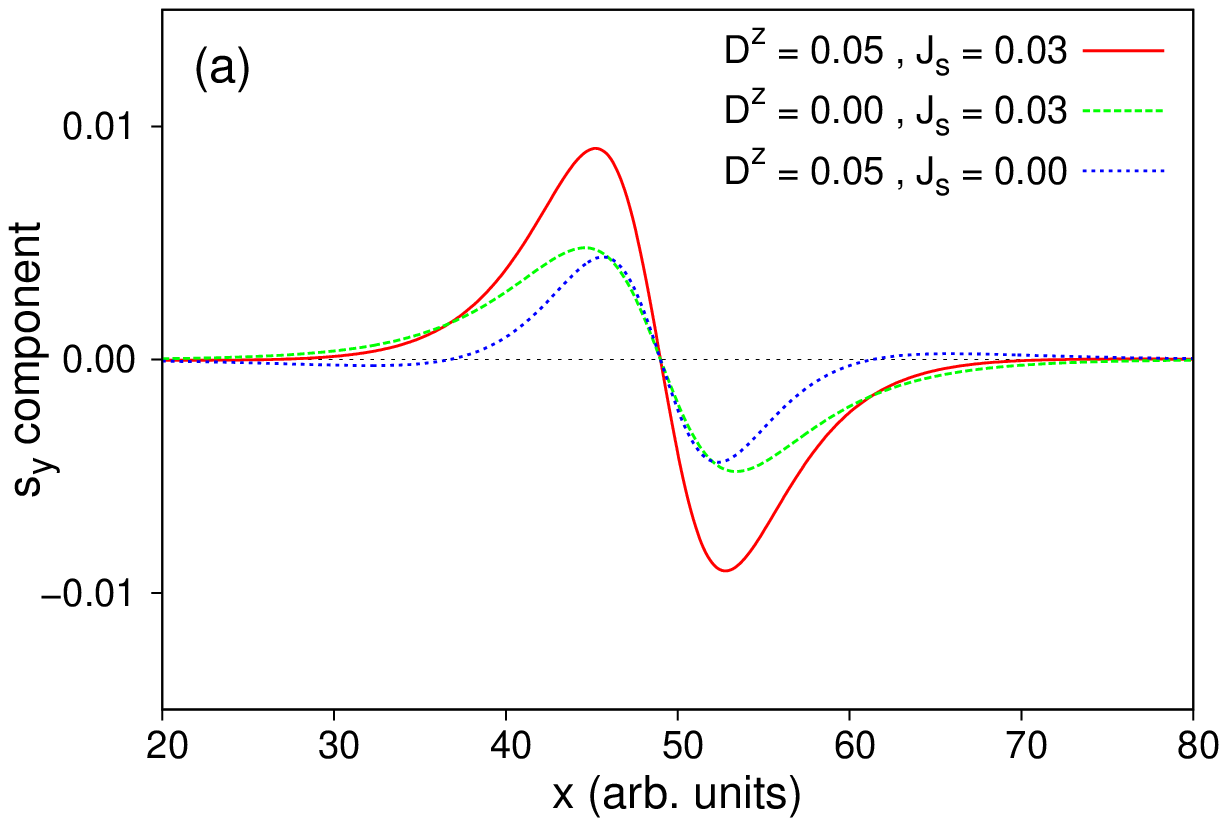}\\
\includegraphics[width=1.0\columnwidth]{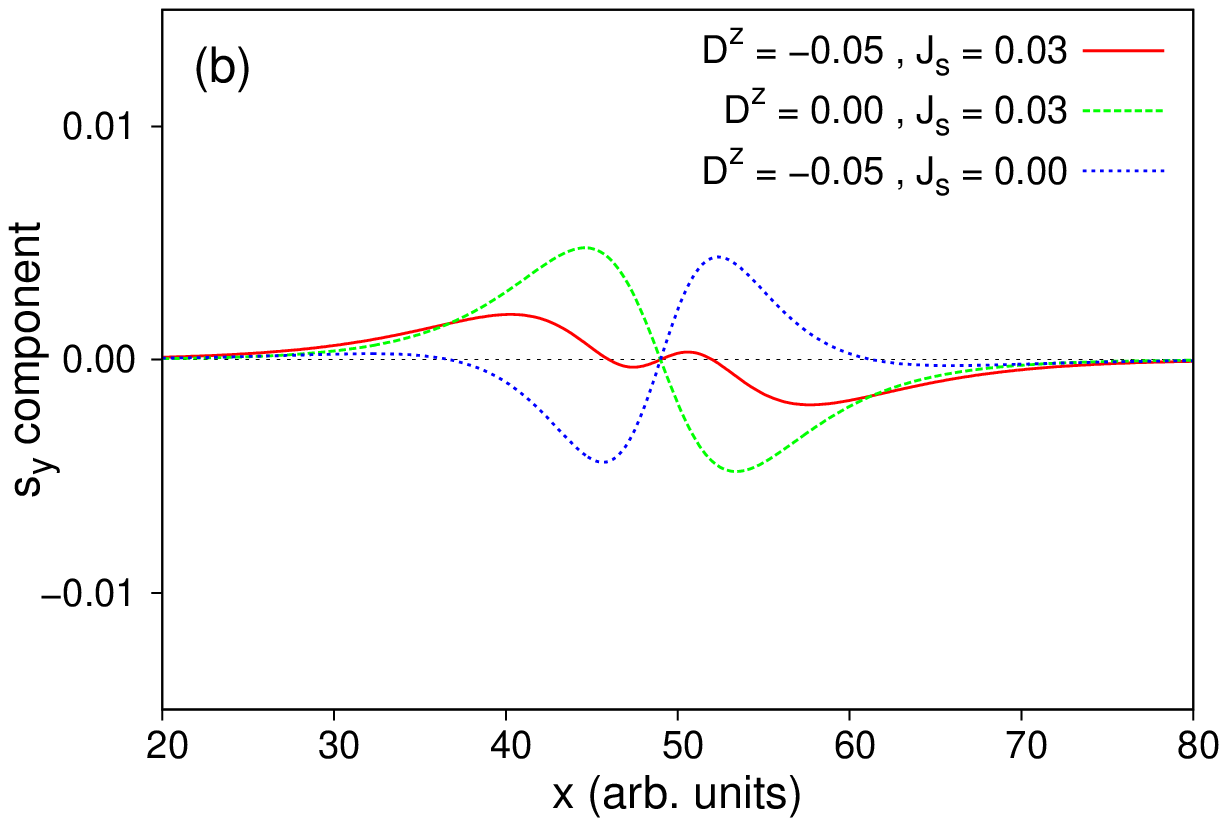}\\
\caption{(Color online) Twisting of DWs obtained from the continuum model. The dimensionless interaction parameters are $\mathcal{A}=1, \mathcal{K}=-0.05, \mathcal{D}=-0.1$, and the $\mathcal{D}^{z}$ and $\mathcal{J}_{\textrm{s}}$ values given in the labels. The distance $x$ is given in units determined by the parameter values and the functional Eq.~(\ref{functional}). The lineshape of the twisting strongly depends on the relative sign of $\mathcal{D}^{z}$ and $\mathcal{J}_{\textrm{s}}$.}
\label{contmodel}
\end{figure}

We have determined the equilibrium domain wall profile by rewriting Eq.~(\ref{functional}) into spherical coordinates for the spin field, and numerically solving the Euler--Lagrange equations with the boundary conditions corresponding to a N\'{e}el DW -- see Appendix~\ref{secS2} for the derivation. The twisting obtained from the numerical solution for specific parameter sets is illustrated in Fig.~\ref{contmodel}, by using dimensionless ferromagnetic coupling $\mathcal{A}=1$, easy-axis anisotropy $\mathcal{K}=-0.05$, and DM interaction $\mathcal{D}=-0.1$, the latter being responsible for fixing the right-handed N\'{e}el rotation of the DW observed in the spin dynamics simulations for Co/Ir$_{n}$/Pt(111). In the absence of the $\mathcal{D}^{z}$ and $\mathcal{J}_{\textrm{s}}$ terms, it is known that all domain wall profiles are equivalent under an appropriate rescaling of the length unit\cite{Heide-PRB}. This is no longer the case here; it can be seen in Fig.~\ref{contmodel} that if $\mathcal{J}_{\textrm{s}}$ is finite, then the $s_{y}$ component changes sign in the middle of the DW, while for $\mathcal{D}^{z}$ further sign changes may occur away from the center. If the sign of $\mathcal{J}_{\textrm{s}}$ and $\mathcal{D}^{z}$ is the same (Fig.~\ref{contmodel}(a) solid curve), the two types of twisting add up, for a net effect that is similar to the one observed for Co/Ir$_{4}$/Pt(111) in Fig.~\ref{deviation}. For different signs (Fig.~\ref{contmodel}(b) solid curve), it is possible that the twisting almost disappears around the middle of the DW, similarly to the case of  Co/Ir$_{3}$/Pt(111) in Fig.~\ref{deviation}. It should be noted that further DW twisting shapes may be obtained by modifying the ratio of $\mathcal{J}_{\textrm{s}}$ and $\mathcal{D}^{z}$ besides their sign.

\section{Conclusion}

In summary, we examined the Co/Ir$_{n}$/Pt$(111)$ ($n=0,\dots,6$) ultrathin films by a combined approach of first-principles calculations and spin dynamics simulations. We determined the Co-Co magnetic exchange interaction tensors between different pairs of neighbors and the magnetic anisotropies for FCC and HCP growth of the Co overlayer, depending on the number of the Ir buffer layers. We found strong nearest-neighbor ferromagnetic isotropic exchange interactions in the Co layer and an easy-axis out-of-plane anisotropy for the films in FCC geometry, independent from their thickness. Our simulations have proven that the system prefers N\'{e}el walls over Bloch walls and, in agreement with related experiments\cite{PhysRevB.90.020402}, the chirality of the N\'{e}el walls switches from left-handed to right-handed when the Ir layer is inserted between the Co monolayer and the Pt$(111)$ substrate. Both facts were uniquely attributed to the in-plane components of the Dzyaloshinskii-Moriya vectors, emphasizing that nearest-neighbor in-plane DM vectors dominate in the Co/Pt(111) system, whereas for Co/Ir$_n$/Pt(111) the second- and third-neighbor in-plane DM vectors are the largest ones. Furthermore, we have found that the width of domain walls is significantly smaller in the presence of the Ir buffer layers, owing to the decreased ferromagnetic isotropic exchange interactions and the increased magnetic anisotropy energy.

We also demonstrated the existence of a twisted domain wall profile, where the spins are not perfectly coplanar as in the ideal N\'{e}el wall. This effect was attributed to the out-of-plane component of the Dzyaloshinskii-Moriya vectors, which are not forbidden by symmetry in the triangular lattice on the $(111)$ surfaces of cubic lattices, and to the $\vec{J}_{ij,\textrm{s}}$ vectors constructed from the symmetric off-diagonal part of the interaction tensor. Based on symmetry arguments we have shown that the twisting must disappear if the normal vector of the DW is within the mirror plane of the system, but it is present for arbitrarily small values of these specific interaction coefficients if the normal vector is perpendicular to the mirror plane. We managed to qualitatively reproduce the different twisting lineshapes observed for different numbers of Ir buffer layers by constructing an appropriate micromagnetic model containing the out-of-plane component of the Dzyaloshinskii-Moriya vectors and the $\vec{J}_{ij,\textrm{s}}$ vectors.

\begin{acknowledgments}
Financial support of the National Research, Development and Innovation Office of Hungary under Projects No.\ K115575 and No.\ PD120917 is gratefully acknowledged. K.P. acknowledges the SASPRO Fellowship of the Slovak Academy of Sciences (Project No.\ 1239/02/01).
\end{acknowledgments}

\appendix

\section{Construction of the continuum model\label{secS1}}

In this Appendix we derive the energy functional Eq.~(\ref{functional}) used for the description of twisted DWs by starting from the atomistic model. First we consider the exchange interaction tensor $\mathbf{J}_{ij}$ between two NNs displaced along the $x$ axis. In the $C_{3\textrm{v}}$ symmetry class of the system, mirror planes connect next-nearest neighbors, and they go through the center of the line connecting the NNs, see Fig.~\ref{DMvectors}(b). Mirroring the system switches the spins $\vec{s}_{i}$ and $\vec{s}_{j}$ and also transforms them as axial vectors. Due to this symmetry, both the $\vec{D}_{ij}$ and $\vec{J}_{ij,\textrm{s}}$ vectors must lie in the mirror plane. This simplifies the form of the interaction tensor to
\begin{eqnarray}
\mathbf{J}_{ij}=\left[\begin{array}{ccc}J'_{ij}+\Delta J_{ij,\textrm{c}} & D_{ij}^{z} & -D_{ij}^{y} \\ -D_{ij}^{z} & J'_{ij} & J_{ij,\textrm{s}}^{y} \\ D_{ij}^{y} & J_{ij,\textrm{s}}^{y} & J'_{ij}+\Delta J_{ij}^{zz}\end{array}\right],\label{eqnS1}
\end{eqnarray}
which has 6 independent components. The possible rotations do not decrease the number of independent components further. The interaction tensors with the other 5 nearest neighbors can be obtained by performing the necessary symmetry operations.

For constructing the continuum model, one has to replace the spin vectors $\vec{s}_{i}$ by the field $\vec{s}\left(\vec{R}_{i}\right)$, expand the spins at the neighboring lattice sites in Taylor series,
\begin{equation}
\vec{s}\left(\vec{R}_{i}+\vec{\delta}\right)=\sum_{n=0}^{\infty}\frac{1}{n!}\left(\vec{\delta}\cdot\vec{\nabla}\right)^{n}\vec{s}\left(\vec{R}_{i}\right),
\end{equation}
then perform the summation over the NNs. For every independent component in Eq.~(\ref{eqnS1}), we truncated the Taylor series at the first nonvanishing finite derivative. This leads to the energy expression
\begin{eqnarray}
E=& \int  \left( \tilde{w}_{\mathcal{A}}+\tilde{w}_{\mathcal{K}}+\tilde{w}_{\mathcal{D}}+\tilde{w}_{\mathcal{D}^{z}}+\tilde{w}_{\mathcal{J}_{\textrm{s}}} \right)\textrm{d}^{2}\vec{r},\label{eqnS3}
\end{eqnarray}
where the $\tilde{w}$ notation denotes that the energy densities are expressed for a two-dimensional system. The final form of Eq.~(\ref{functional}) is obtained after simplifying Eq.~(\ref{eqnS3}) to one spatial dimension, where it is assumed that the spins are parallel when the integration is performed along the $y$ direction.

The first term in Eq.~(\ref{eqnS3}), corresponding to the exchange stiffness, reads
\begin{eqnarray}
\tilde{w}_{\mathcal{A}}=\tilde{A}\left(\vec{\nabla}\vec{s}\right)^{2},
\end{eqnarray}
which is obtained from the isotropic exchange interactions $J_{ij}=\frac{1}{3} \textrm{Tr}\mathbf{J}_{ij}$ or the related coefficient $J'_{ij}=J_{ij}-\frac{1}{3}\Delta J_{ij,\textrm{c}}-\frac{1}{3}\Delta J_{ij}^{zz}$ in Eq.~(\ref{eqnS1}). Note that $\tilde{A}>0$ denotes ferromagnetic coupling in this expression.

The anisotropy term is
\begin{eqnarray}
\tilde{w}_{\mathcal{K}}=\tilde{\mathcal{K}}s_{z}^{2},
\end{eqnarray}
which contains contributions from the on-site anisotropy term Eq.~(\ref{eqn4}), as well as the leading-order corrections from the coefficients $\Delta J_{ij}^{zz}$ and $\Delta J_{ij,\textrm{c}}$; see the two-site magnetic anisotropy $\Delta J$ defined in Sec.~\ref{sec2}. We mention that $\Delta J_{ij,\textrm{c}}$, known as the compass anisotropy\cite{RevModPhys.87.1}, also leads to a term that prefers Bloch DWs for $\Delta J_{ij,\textrm{c}}<0$ when it is expanded up to second-order spatial derivatives. This is analogous to the role of the magnetostatic dipolar interaction; however, as it was discussed in Sec.~\ref{sec3b} the rotation of the DW from N\'{e}el-type towards Bloch-type is a threshold effect, and we have not observed it during the simulations. Therefore, we have not included
the term related to $\Delta J_{ij,\textrm{c}}$ in the energy functionals Eq.~(\ref{functional}) and Eq.~(\ref{eqnS1}), used for the description of the DW twisting.

The DM interaction or linear Lifshitz invariant reads
\begin{eqnarray}
\tilde{w}_{\mathcal{D}}=\tilde{\mathcal{D}}\left(s_{z}\partial_{x}s_{x}-s_{x}\partial_{x}s_{z}+s_{z}\partial_{y}s_{y}-s_{y}\partial_{y}s_{z}\right),\label{eqnS6}
\end{eqnarray}
and it is obtained from the in-plane components of the DM vectors $D_{ij}^{y}$.

Since Eq.~(\ref{eqnS6}) is the only expression containing first-order derivatives allowed in the $C_{n\textrm{v}}$ symmetry class\cite{Bogdanov3}, the out-of-plane component of the DM vectors $D_{ij}^{z}$ may only show up in the form of higher-order derivatives. The leading contribution is
\begin{align}
\tilde{w}_{\mathcal{D}^{z}}=\tilde{\mathcal{D}}^{z}\left(s_{x}\left(\partial_{x}^{3}-3\partial_{x}\partial_{y}^{2}\right)s_{y}-s_{y}\left(\partial_{x}^{3}-3\partial_{x}\partial_{y}^{2}\right)s_{x}\right).\label{eqnS7}
\end{align}

The last term in Eq.~(\ref{eqnS3}) reads
\begin{align}
\tilde{w}_{\mathcal{J}_{\textrm{s}}}=&\tilde{\mathcal{J}}_{\textrm{s}}(s_{y}\left(\partial_{x}^{2}-\partial_{y}^{2}\right)s_{z}+s_{z}\left(\partial_{x}^{2}-\partial_{y}^{2}\right)s_{y} \nonumber 
 \\
&+s_{x}\left(2\partial_{x}\partial_{y}\right)s_{z}+s_{z}\left(2\partial_{x}\partial_{y}\right)s_{x}),\label{eqnS8}
\end{align}
which contains the contributions from the interaction coefficients $J_{ij,\textrm{s}}^{y}$ in the discrete model.

The following symmetry argument proves why during the twisting a node of the $y$ spin component appears at the center of the DW. As it can be seen in Fig.~\ref{spindin}, the out-of-plane component of the spins is an odd function of the distance from the center of the wall, corresponding to a $\tanh$ function in the ideal case\cite{Landau}. On the other hand, the in-plane component is an even function, ideally $1/\cosh$. The system may gain energy from $\tilde{w}_{\mathcal{D}^{z}}$ in Eq.~(\ref{eqnS7}) or $\tilde{w}_{\mathcal{J}_{\textrm{s}}}$ in Eq.~(\ref{eqnS8}) if the energy densities are even functions, since integrating over an odd function yields zero. In both cases this means that the $s_{y}$ component must be an odd function, in agreement with the simulational observations. Such a twisting is energetically preferable for an arbitrarily small value of these interaction coefficients, in contrast to the rotation towards the direction of the Bloch-type DW; this can be demonstrated by constructing the Euler--Lagrange equations (\ref{eqnS9})-(\ref{eqnS10}) given in Appendix~\ref{secS2} below.

From Eqs.~(\ref{eqnS7})-(\ref{eqnS8}) it can also be seen why the twisting disappears for N\'{e}el-type DWs with normal vector along the $y$ direction. The term $\tilde{w}_{\mathcal{D}^{z}}$ Eq.~(\ref{eqnS7}) exactly cancels when the normal vector of the wall is along the $y$ direction. In the case of $\tilde{w}_{\mathcal{J}_{\textrm{s}}}$, it will still contain only the $s_{y}$ and $s_{z}$ spin components as in Eq.~(\ref{functional}). Consequently, it can only induce a twisting if originally the spins in the domain wall  lie in the $xz$ plane, which corresponds to a Bloch DW with normal vector along the $y$ direction. For completeness, we mention that $\tilde{w}_{\mathcal{D}^{z}}$ also induces a twisting for Bloch DWs with normal vector along the $x$ direction, but the $\tilde{w}_{\mathcal{J}_{\textrm{s}}}$ term only induces a twisting for N\'{e}el-type DWs oriented in this direction; for a summary see Table~\ref{tableS1}.

\begin{table}
  \centering
  \begin{ruledtabular}
    \begin{tabular}{ccccc}
          & \multicolumn{2}{c}{$\tilde{w}_{\mathcal{D}^{z}}$} & \multicolumn{2}{c}{$\tilde{w}_{\mathcal{J}_{\textrm{s}}}$} \\
    normal vector & Bloch wall & N\'{e}el wall & Bloch wall & N\'{e}el wall \\
    \hline
    $x$ & yes   & yes   & no    & yes \\
    $y$ & no    & no    & yes   & no \\
    \end{tabular}%
  \end{ruledtabular}
\caption{Table summarizing for which type and normal vector of the DW the interaction terms $\tilde{w}_{\mathcal{D}^{z}}$ and $\tilde{w}_{\mathcal{J}_{\textrm{s}}}$ can induce a twisting. The first term $\tilde{w}_{\mathcal{D}^{z}}$ does not induce a twisting when the normal vector is along the $y$ direction, while for the second term $\tilde{w}_{\mathcal{J}_{\textrm{s}}}$ this is forbidden by symmetry if the spins in the DW lie in the $yz$ plane.\label{tableS1}}%
\end{table}%

\section{Euler--Lagrange equations\label{secS2}}

In order to determine the domain wall profile from Eq.~(\ref{functional}), we represented the spin field in spherical coordinates,
\begin{equation}
\vec{s}=\left[\begin{array}{c}\sin\vartheta\cos\varphi \\ \sin\vartheta\sin\varphi \\ \cos\vartheta\end{array}\right],
\end{equation}
where the different energy contributions may be expressed as
\begin{align}
w_{\mathcal{A}}=&\mathcal{A}  \left(\dot{\vartheta}^{2} +  \dot{\varphi}^{2}  \sin^{2} \vartheta  \right),
\end{align}
\begin{align}
w_{\mathcal{J}_{\textrm{s}}}=&-\mathcal{J}_{\textrm{s}} \left(\dot{\vartheta}^{2}\cos\vartheta\sin\vartheta\sin\varphi + \dot{\vartheta}\dot{\varphi}\sin^{2}\vartheta\cos\varphi \right),
\end{align}
\begin{align}
w_{\mathcal{K}}=&\mathcal{K} \cos^{2} \vartheta,
\end{align}
\begin{align}
w_{\mathcal{D}}=&\mathcal{D} \left( \dot{\vartheta} \cos \varphi  -  \dot{\varphi}  \sin \vartheta\cos\vartheta \sin \varphi \right) ,
\end{align}
\begin{align}
w_{\mathcal{D}^{z}}=&\mathcal{D}^{z} \Big(  3\ddot{\vartheta} \dot{\varphi} \cos\vartheta\sin\vartheta - 3\dot{\vartheta}^{2}\dot{\varphi}\sin^{2}\vartheta \nonumber \\
&+ 3\dot{\vartheta}\ddot{\varphi}\cos\vartheta\sin\vartheta   - \dot{\varphi}^{3}\sin^{2}\vartheta+ \dddot{\varphi}\sin^{2}\vartheta   \Big).
\end{align}

The equilibrium domain wall profile can be determined by solving the Euler--Lagrange equations corresponding to Eq.~(\ref{functional}) using the general formulas
\begin{align}
\sum_{n=0}^{\infty}\left(-\frac{\textrm{d}\:}{\textrm{d}x}\right)^{n}\frac{\partial\:\:\:\:\:\:}{\partial \vartheta^{(n)}}\left(w_{\mathcal{A}}+w_{\mathcal{J}_{\textrm{s}}}+w_{\mathcal{K}}+w_{\mathcal{D}}+w_{\mathcal{D}^{z}}\right)=&0,
\\
\sum_{n=0}^{\infty}\left(-\frac{\textrm{d}\:}{\textrm{d}x}\right)^{n}\frac{\partial\:\:\:\:\:\:}{\partial \varphi^{(n)}}\left(w_{\mathcal{A}}+w_{\mathcal{J}_{\textrm{s}}}+w_{\mathcal{K}}+w_{\mathcal{D}}+w_{\mathcal{D}^{z}}\right)=&0,
\end{align}
appropriate for higher-order derivatives. This yields

\begin{widetext}

\begin{align}
&\mathcal{A} \left(-  2\ddot{\vartheta}+2\dot{\varphi}^{2}\cos\vartheta\sin\vartheta  \right) +   \mathcal{D} 2\dot{\varphi} \sin^{2}\vartheta\sin\varphi-\mathcal{K}2\sin\vartheta\cos\vartheta &  \nonumber \\
&+\mathcal{D}^{z} \left( 6 \ddot{\vartheta}  \dot{\varphi}\cos^{2}\vartheta - 6  \dot{\vartheta}^{2} \dot{\varphi}  \sin\vartheta \cos\vartheta + 6 \dot{\vartheta} \ddot{\varphi}  \cos^{2}\vartheta - 2 \dot{\varphi}^{3} \sin\vartheta \cos\vartheta +  2\dddot{\varphi} \sin\vartheta\cos\vartheta \right) & \nonumber \\
& +\mathcal{J}_{\textrm{s}}\left(2\ddot{\vartheta}\cos\vartheta\sin\vartheta\sin\varphi+ \dot{\vartheta}^{2}\left(\cos^{2}\vartheta-\sin^{2}\vartheta\right)\sin\varphi +2\dot{\vartheta}\dot{\varphi}\sin\vartheta\cos\vartheta\cos\varphi-\dot\varphi^{2}\sin^{2}\vartheta\sin\varphi +\ddot{\varphi}\sin^{2}\vartheta\cos\varphi\right)= 0,\label{eqnS9}
\\
&\mathcal{A} \left( -4\dot{\vartheta}\dot{\varphi}\sin\vartheta\cos\vartheta- 2\sin^{2}\vartheta \ddot{\varphi}    \right) - \mathcal{D}2\dot{\vartheta}\sin\varphi\sin^{2}\vartheta+\mathcal{J}_{\textrm{s}}\left(\ddot\vartheta\sin^{2}\vartheta\cos\varphi +\dot\vartheta^{2}\sin\vartheta\cos\vartheta\cos\varphi\right) & \nonumber \\
&+ \mathcal{D}^{z} \left( -2\dddot{\vartheta}\sin\vartheta\cos\vartheta + 6\ddot{\vartheta}\dot{\vartheta}\sin^{2}\vartheta+    2\dot{\vartheta}^{3}\sin\vartheta\cos\vartheta + 6\dot{\vartheta}\dot{\varphi}^{2}\sin\vartheta\cos\vartheta + 6\dot{\varphi}\ddot{\varphi}\sin^{2}{\vartheta}    \right)  =0 .\label{eqnS10}&
\end{align}

\end{widetext}

The Euler--Lagrange equations were solved with the boundary conditions describing the right-rotating cycloidal N\'{e}el domain wall observed in the simulations in the presence of the Ir buffer layers, see Fig.~\ref{spindin}(b). These correspond to $\vartheta=\pi, \varphi=\pi$ as $x\rightarrow-\infty$ and $\vartheta=0, \varphi=\pi$ as $x\rightarrow\infty$. By looking at the Euler--Lagrange equations it can clearly be seen that the perfect N\'{e}el shape $\varphi\equiv\pi$ cannot be an equilibrium solution for any finite value of $\mathcal{D}^{z}$ or $\mathcal{J}_{\textrm{s}}$, and a twisting will occur.

\bibliographystyle{apsrev4-1}
\bibliography{Co1IrnPt111}
\end{document}